\documentclass[aps,prl,twocolumn,amsmath,amssymb]{revtex4}

\usepackage{graphicx}
\usepackage{dcolumn}
\usepackage{bm}

\newcommand{\tal}{TlCuCl$_3$}
\newcommand{\kal}{KCuCl$_3$}
\newcommand{\nal}{NH$_4$CuCl$_3$}

\begin{document}

\preprint{APS/123-QED}

\title{Magnetoelastic Coupling in the Spin-Dimer System TlCuCl$_3$}

\author{N. Johannsen$^1$, A. Vasiliev$^2$, A. Oosawa$^3$, H. Tanaka$^4$, and T. Lorenz$^1$}
\affiliation{ $^1$II.\,Physikalisches Institut, Universit\"{a}t zu
K\"{o}ln,Z\"{u}lpicher Str. 77, 50937 K\"{o}ln, Germany \\
$^2$Department of Low Temperature Physics, Moscow State
University, Moscow 119992, Russia \\
$^3$Advanced Science Research Center, Japan Atomic Energy Research
Institute, Tokai, Ibaraki 319-1195, Japan \\
$^4$Dep. of Physics, Tokyo Inst. of Technology, Oh-okayama,
Meguro-ku, Tokyo 152-8551, Japan}

\date{\today}

\begin{abstract}
We present high-resolution measurements of the thermal expansion
and the magnetostriction of TlCuCl$_{3}$ which shows
field-induced antiferromagnetic order. We find pronounced
anomalies in the field and temperature dependence of different
directions of the lattice signaling a large magnetoelastic
coupling. The phase boundary is extremely sensitive to pressure,
e.g.\ the transition field would change by about $\pm 185$~\%/GPa
under uniaxial pressure applied along certain directions. This
drastic effect can unambiguously be traced back to changes of the
intradimer coupling under uniaxial pressure. The interdimer
couplings remain essentially unchanged under pressure, but
strongly change when Tl is replaced by K.

\end{abstract}

\maketitle

One of the most simple quantum spin system is a spin-1/2 dimer.
If such dimers are weakly coupled to each other, very rich and
fascinating physical properties are predicted for various
theoretical models and can be observed experimentally in suitable
materials. For example, the two-dimensional Shastry-Sutherland
model\,\cite{shast81b} is realized experimentally by
SrCu$_2$(BO$_3$)$_2$ and its low-temperature magnetization as a
function of magnetic field shows distinct plateaus at certain
fractional values of the saturation
magnetization\,\cite{kagey99}. Magnetization plateaus are also
observed in the three-dimensional spin-dimer system
NH$_4$CuCl$_3$\,\cite{shiramura98a}. Such plateaus are, however,
absent in its iso-structural (at 300\,K) counterparts $R$CuCl$_3$
with $R={\rm Tl}$ and K, which both have a non-magnetic ground
state up to a certain magnetic field\,\cite{Shiramura1997}. Above
this field a N\'{e}el order with staggered magnetization
perpendicular to the applied field
occurs\,\cite{tanaka2001,Oosawa2002d} and it has been proposed
that this transition should be viewed as a Bose-Einstein
condensation (BEC) of
magnons\,\cite{Nikuni2000,matsumoto2002,ruegg03,matsumoto2004}.
According to a recent neutron scattering study, the different
behavior of \nal\ is connected with two structural phase
transitions in that compound\,\cite{ruegg04a}. Despite their
qualitative similarity, the magnetic systems of \tal\ and \kal\
show pronounced quantitative differences. The triplet excitations
of \tal\ are strongly dispersive, whereas those of \kal\ have a
weak dispersion\,\cite{matsumoto2004,cavadini2001,cavadini2000}.
Consequently, the minimum gap $\Delta \simeq 8$\,K is
significantly smaller than the intradimer coupling $J\simeq 64$\,K
for \tal , whereas this difference is much weaker for \kal\
($\Delta \simeq 30$\,K and $J\simeq
50$\,K)\,\cite{matsumoto2004,cavadini2001,cavadini2000}. The very
different behavior of the $R$CuCl$_3$ series shows that small
structural differences strongly influence the magnetic subsystem.
Evidence for a strong magnetoelastic coupling in \tal\ is also
found in ultrasound and NMR data, which indicate that the phase
transition of \tal\ has a significant contribution of first-order
character\,\cite{Sherman2003,Vyaselev2004}. A BEC is expected to
be of second order, but spin-phonon coupling can drive a
continuous transition into a first-order phase
transition\,\cite{ito1978}. Moreover, hydrostatic pressure of
less than 0.5~GPa is already sufficient to close $\Delta$ and to
induce antiferromagnetic order in \tal\ without magnetic
field\,\cite{Oosawa2003a,tanaka2003a}. A microscopic
understanding of the relevant changes under pressure is still
missing. In particular, it is not clear why external pressure
decreases $\Delta$, whereas the substitution of the Tl$^+$ ions
by the smaller K$^+$ ions increases $\Delta$.

We present high-resolution measurements of the thermal expansion
and the magnetostriction of a single crystal of TlCuCl$_{3}$.
Using a capacitance dilatometer we studied the length changes
perpendicular to the $(010)$ and $(10\overline{2})$ cleavage
planes of the monoclinic crystal structure\,\cite{tanaka2001}.
Via thermodynamic relations we derive the uniaxial pressure
dependencies of the transition temperatures $T_c$, of the
transition fields $H_c$, of the spin gap $\Delta$, and of the
magnetic coupling constants. For $\Delta$ we find huge pressure
dependencies of about $\pm 185$\,\%/GPa for uniaxial pressure
perpendicular to the $(010)$ or $(10\overline{2})$ planes,
respectively. The uniaxial pressure dependencies of $\Delta$
unambiguously correlate with changes of the intradimer coupling
$J$ under pressure. In contrast to recent
assumptions\,\cite{matsumoto2004}, pressure-dependent changes of
the interdimer coupling $J'$ play a minor role. Thus, the weaker
$J'$ of \kal\ is not a consequence of chemical pressure. This
gives clear evidence that the Tl$^+$ and K$^+$ ions are directly
involved in the superexchange which is responsible for the
relevant interdimer coupling and experimentally confirms the
theoretical result of a significantly stronger superexchange via
Tl$^+$ than via K$^+$ for $R$CuCl$_3$\,\cite{saha-dasgupta2002}.

\begin{figure}
\includegraphics[angle=0,width=8cm,clip]{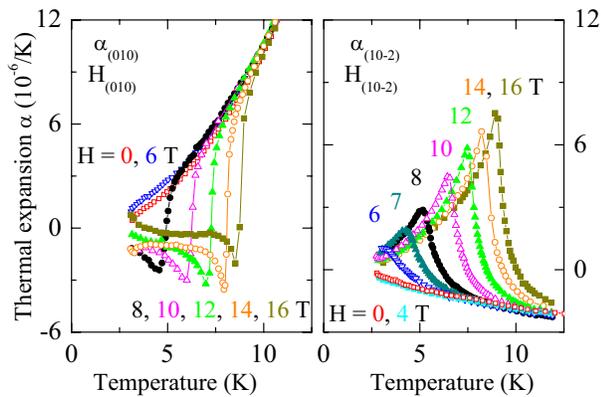}
\caption{\label{alpha} Thermal expansion $\alpha_i$ perpendicular
to the $(010)$ and $(10\overline{2})$ planes for different
magnetic-field strengths applied parallel to the respective
measurement direction.}
\end{figure}

In Fig.~\ref{alpha} we show the longitudinal thermal expansion
$\alpha_i $ along different directions $i$ for various values of
a magnetic field applied parallel to the respective measurement
directions. In low fields there are no anomalies of $\alpha_i$,
whereas above $6$~T strong anomalies of opposite signs appear for
both directions. With increasing field the anomalies increase and
systematically shift to higher temperature. These anomalies
signal spontaneous distortions below the phase transition: a
spontaneous elongation perpendicular to the $(010)$ and a
spontaneous contraction perpendicular to the $(10\overline{2})$
plane.

\begin{figure}
\includegraphics[angle=0,width=7.5cm,clip]{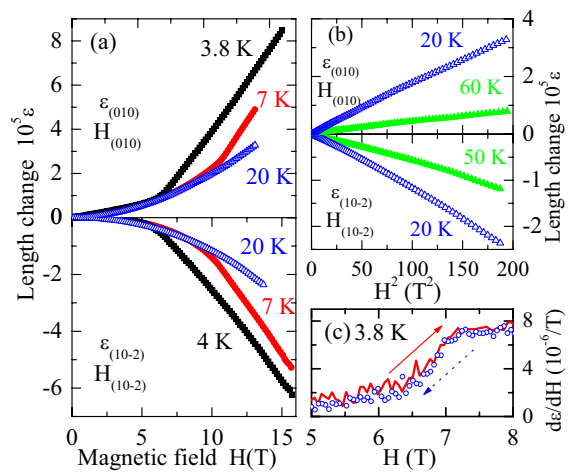}
\caption{\label{MS} Magnetostriction  $\epsilon_i=\Delta L_i/L_i$
perpendicular to the $(010)$ and the $(10\overline{2})$ planes at
selected temperatures. Panel (a) shows  $\epsilon_i \, vs.\, H$
and (b) $\epsilon_i \,vs.\, H^2$, respectively. Panel (c)
displays the field-derivatives $\frac{\partial
\epsilon_i}{\partial H}$ for $i=(010)$ obtained with increasing
(line) and decreasing (symbols) field at $T=3.8$\,K. }
\end{figure}

Fig.~\ref{MS} displays representative magnetostriction
measurements at different constant temperatures for both
measurement directions. The relative length changes
$\epsilon_i=\Delta L_i(H)/L_i$ as a function of field for
$i=(010)$ and $(10\overline{2})$ are again of comparable size but
of opposite sign. The phase transition causes a sharp kink in
$\epsilon_i$ as a function of field. With increasing temperature
these kinks shift to higher fields and cannot be observed anymore
in the studied field range above about 9\,K. For $H>H_c$,
$\epsilon_i$ changes essentially linear with field, whereas for
smaller fields or for higher temperatures $\epsilon_i$ is
proportional to $H^2$ (see Fig.\,\ref{MS}b). A kink in
$\epsilon_i$ is typical for a second-order phase transition,
which should give rise to a jump-like anomaly in the
field-derivative $\partial \epsilon_i/\partial H$. As mentioned
above, there is some indication for a first-order contribution to
the phase transition in \tal . A small region of coexisting
phases around $H_c$ as proposed from the NMR
data\,\cite{Vyaselev2004} can be neither confirmed nor ruled out
by our measurements of the macroscopic length changes. We can,
however, exclude that there is a significant hysteresis of about
$0.5$~T between the $H_c$ values obtained with increasing and
decreasing magnetic field as has been observed in an ultrasound
study\,\cite{Sherman2003}. In Fig.\,\ref{MS}c we compare $\partial
\epsilon_i/\partial H$ obtained with increasing and decreasing
magnetic field. As expected for a second-order phase transition
there is a (broadened) jump at $H_c$ and both curves agree well
with each other over the entire field range. Thus, any hysteresis
of $H_c$ is restricted to less than our field resolution of about
50\,mT.

\begin{figure}
\includegraphics[angle=0,width=7.2cm, clip]{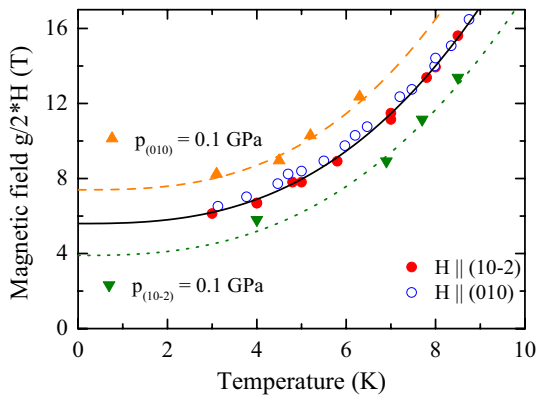}
\caption{\label{phadi} Phase diagram of \tal\ from the
magnetostriction and thermal expansion measurements perpendicular
to the $(010)$ ($\circ$) and $(10\overline{2})$ ($\bullet$) planes
normalized by the respective $g$ factors\,\cite{Oosawa2001}. The
triangles display the hypothetical shift of selected $T_c$ and
$H_c$ values for a uniaxial pressure of 0.1\,GPa perpendicular to
$(010)$ ($\blacktriangle$) and $(10\overline{2})$
($\blacktriangledown$) calculated by Eq.\,\ref{ehrenfest}. The
lines are power-law fits of the phase
boundaries.\label{phadi_druck_beide}}
\end{figure}

In Fig.~\ref{phadi} we show the phase diagram obtained from our
data (circles) together with a power-law fit (solid line) of the
form $(g/2)[H_c(T)-H_c(0)]\propto T^\Phi$. This fit yields
$\Phi=2.6$ and $H_c(0)=5.6$\,T. The exponent is larger than the
value of 2.1 obtained in Ref.\,\cite{Oosawa2001} (for $T<4$\,K),
but agrees well with the result of Quantum Monte Carlo (QMC)
simulations\,\cite{wessel2001}. According to a more recent QMC
study\,\cite{nohadani2004}, $\Phi $ sensitively depends on the
temperature range of the fit and in the low-temperature limit
$\Phi=1.5$ is approached, which agrees to the expected value for
a BEC. Thus, our larger value of $\Phi$ arises most probably from
the used temperature range (3\,K$<T<9$\,K), but one should also
keep in mind that $\Phi$ could change due to the finite
spin-phonon coupling, which is not considered in the
models\,\cite{Nikuni2000,matsumoto2002,wessel2001,nohadani2004}.

The anomalies at the phase boundary allow to derive the uniaxial
pressure dependencies of $T_c$ and $H_c$ by the Ehrenfest
relations
\begin{equation}
\frac{\partial T_c}{\partial
p_{i}}=V_{m}\,T_c\,\frac{\Delta\alpha_{i}}{\Delta C_p}\,\, \mbox{
and }\,\,
 \frac{\partial H_c}{\partial p_{i}}
 = V_{m}\frac{\Delta \frac{\partial \epsilon_i}{\partial
H}}{\Delta \frac{\partial M_{mol}}{\partial H}}\,.
\label{ehrenfest}
\end{equation}
Here, $V_m$ is the molar volume, $\Delta \alpha_{i}$ is the height
of the thermal-expansion anomaly (see Fig.~\ref{alpha}) and
${\Delta C_p}$ that of the corresponding specific-heat
anomaly\,\cite{Oosawa2001}, $\Delta \frac{\partial
\epsilon_i}{\partial H}$ is the slope change of $\epsilon_i$ at
$H_c$ (see Fig.~\ref{MS}) and $\Delta \frac{\partial
M_{mol}}{\partial H}$ the corresponding slope change of the
magnetization\,\cite{Shiramura1997}.

With $C_p$ and $M$ from Refs.\,\cite{Oosawa2001,Shiramura1997} we
find huge uniaxial pressure dependencies of $T_c$ and $H_c$,
e.\,g.\ $\frac{\partial T_c}{\partial p_{010}}\simeq -9$\,K/GPa
for $T_c=7.2$\,K and $H=12$\,T, or $\frac{\partial H_c}{\partial
p_{10\overline{2}}}\simeq -8$\,T/GPa for $H_c=6$\,T and $T=4$\,K.
A hypothetical uniaxial pressure of 0.1\,GPa on $(010)$ would
strongly shift the phase boundary towards higher $H_c$ and lower
$T_c$ values, respectively, whereas uniaxial pressure on
$(10\overline{2})$ would cause a shift in the opposite direction
as shown in Fig.~\ref{phadi} by the upward and downward
triangles, respectively. The dashed and dotted lines represent
power-law fits (keeping $\Phi$ fixed) of these hypothetical phase
boundaries under uniaxial pressure, and their extrapolations to
$T=0$\,K reveal the uniaxial pressure dependencies of $\Delta$
\begin{eqnarray*}
\frac{\partial \ln\Delta}{\partial
p_{10\overline{2}}}=-180\frac{\%}{\textnormal{GPa}} \,\,\mbox{ and
} \,\, \frac{\partial \ln\Delta}{\partial
p_{010}}=+190\frac{\%}{\textnormal{GPa}}\,\,.
\end{eqnarray*}
These are huge values, but due to the opposite signs, they almost
cancel each other under hydrostatic pressure. Nevertheless, a
strong decrease of $\Delta $ has been observed under hydrostatic
pressure\,\cite{Oosawa2003a,tanaka2003a}. Thus $\Delta $ should
also strongly decrease for uniaxial pressure along the $[201]$
direction, which is perpendicular to both directions of our
measurements. The geometry of our crystal did not allow
measurements along the $[201]$ direction, but we expect that there
will be similar anomalies at the phase boundary as those
perpendicular to the $(10\overline{2})$ plane.

\begin{figure}
\includegraphics[width=7.2cm, clip]{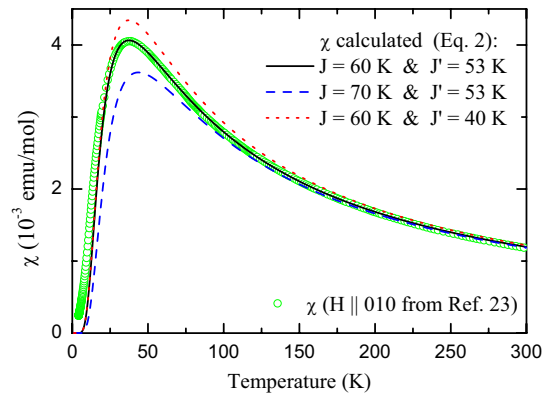}
\caption{\label{chi} Magnetic susceptibility ($\circ $) with a fit
for coupled dimers (solid line, Eq.\,\ref{eq:chiMF}). The dashed
and dotted lines show the expected changes of $\chi $ for an
increase of $J$ and a decrease of $J'$, respectively. Both cases
would cause an increase of the spin gap $\Delta $ (see text.).}
\end{figure}

In a model of dimers coupled by an effective interdimer coupling
$J'$ the magnitude of $\Delta $ is determined by the balance of
$J$ and $J'$\cite{matsumoto2004}. An increase of $J$ will enlarge
$\Delta $, whereas an increase of $J'$ will enhance the bandwidth
of the triplet excitations and therefore lower $\Delta $. Due to
the small value of $\Delta$ compared to $J$ (and $J'$) in \tal\
already moderate pressure-dependent changes of $J$ (or $J'$) may
cause drastic changes of $\Delta $. In order to gain information
whether these changes arise from a pressure-dependence of $J$ or
of $J'$, we fit the magnetic susceptibility
$\chi$\,\cite{Takatsu1997} for temperatures well above the gap by
\begin{eqnarray}
\chi_{MF}(T)=\frac{\chi_0(T)}{1+\chi_0(T)\,J'\,k_B/N_A g^2
\mu_B^2} \label{eq:chiMF}
\,\,\,\,\,\,\,\,\,\,\mbox{ with } \,\, \\
\chi_{0}(T)=\frac{N_A g^2 \mu_B^2 S(S+1)}{3 k_B
T}\frac{2(S+1)\exp(-J/T)}{1+2(S+1)\exp(-J/T)}\,. \label{eq:chi}
\end{eqnarray}
Here, $\chi_0(T)$ is the susceptibility of non-interacting spin
dimers with intradimer coupling $J$, and $\chi_{MF}$ accounts for
a mean-field correction with an effective interdimer coupling
$J'$. As shown by the solid line in Fig.~\ref{chi} the fit for
$T>25$\,K yields a good description of the experimental data for
$J=60$\,K, $J'=53$\,K, and $g=1.48$. Our value of $J$ is close to
the neutron scattering result $J\simeq
64$\,K\,\cite{matsumoto2002,matsumoto2004,cavadini2001,cavadini2000},
whereas our $J'$ is significantly larger than the largest
interdimer coupling ($37$\,K) and our $g$ factor is significantly
smaller than $g_{010} = 2.06$ obtained by ESR\,\cite{Oosawa2001}.
One has to expect quantitative discrepancies due to our
oversimplified model. However, this does hardly affect the
following analysis of the relative variations around the maximum
of $\chi(T)$ arising from pressure-dependent changes of $J$ or
$J'$, because the main result is obtained from the signs of the
uniaxial pressure dependencies of $\chi$ and $\Delta$,
respectively.

Let us discuss the case that $\Delta$ increases as it would under
uniaxial pressure on (010). This may result either from an
increase of $J$ or from a decrease of $J'$. For both cases
$\chi(T)$ can be modeled by Eq.\,\ref{eq:chiMF}. As shown in
Fig.~\ref{chi} the maximum $\chi_{max}$ {\em decreases} if $J$
increases (dashed line) whereas $\chi_{max}$ {\em increases} if
$J'$ decreases (dotted line). Thus, the sign of the uniaxial
pressure dependence of $\chi_{max}$ allows an unambiguous
decision whether the uniaxial pressure dependence of $\Delta$
results from a change of $J$ or of $J'$. Measurements of $\chi $
under uniaxial pressure do not exist. However, the uniaxial
pressure dependence of $\chi$ is related to the magnetostriction
by a Maxwell relation, and $\epsilon_i \propto H^2$ is expected
for a paramagnetic material with $M=\chi H$, i.e.
\begin{equation}
\frac{\partial\epsilon_i}{\partial H}=-\frac{\partial M}{\partial
p_{i}} \,\, \mbox{ and }\,\,
\epsilon_i=-\frac{1}{2}\frac{\partial\chi}{\partial p_{i}}H^2
\,\, .
\end{equation}
As shown in Fig.~\ref{MS}c, the relation $\epsilon_i \propto H^2$
is indeed fulfilled and for the discussed case one finds
$\epsilon_{010}>0$ (upper panel), which implies
$\partial\chi_{max}/\partial p_{010}<0$. A decreasing
$\chi_{max}$ means that the intradimer coupling $J$ increases
(see Fig.\,\ref{chi}). The same argumentation with inverted signs
for all the uniaxial pressure dependencies is valid for pressure
on $(10\overline{2})$. Thus, $J$ is the relevant parameter which
changes under pressure!

The anisotropy of $\frac{\partial \chi_{max}}{\partial p_{i}}$
for $i=(010)$ and $(10\overline{2})$ agrees well with that of
$\frac{\partial \Delta}{\partial p_{i}}$.  This indicates that
$J'$ hardly changes under pressure\,\cite{accident}. Setting
$\frac{\partial J'}{\partial p_{i}}=0$, our model yields
$\frac{\partial \ln \chi_{max}}{\partial p_i}\simeq
-0.7\,\frac{\partial \ln J}{\partial p_i}$ and allows us to
estimate
\begin{eqnarray*}
\frac{\partial \ln J}{\partial p_{10\overline{2}}} \simeq
-34\frac{\%}{\textnormal{GPa}} \,\,\mbox{ and } \,\,
\frac{\partial \ln J}{\partial p_{010}} \simeq
+39\frac{\%}{\textnormal{GPa}}\,\,.
\end{eqnarray*}
The relative changes of $J$ are much smaller than those of
$\Delta$, but because $\Delta $ is much smaller than $J$, the
absolute changes of $J$ and $\Delta $ are not too different ($\pm
22$\,K/GPa and $\pm 14$\,K/GPa, respectively). This can be
interpreted as follows. The pressure-induced change of $J$ causes
mainly a shift of the center of mass of the triplet excitations,
but hardly changes its bandwidth. This is completely different
from what is observed when Tl is substituted by K: The triplet
excitations of \kal\ have a much smaller bandwidth than those of
\tal . Our analysis of the pressure dependencies clearly shows
that this smaller bandwidth is not a consequence of chemical
pressure, although K$^+$ is significantly smaller than Tl$^+$.
Thus the very different values of $J'$ in \kal\ and \tal\ mean
that the K$^+$ and Tl$^+$ ions directly influence the effective
interdimer coupling. This conclusion has been proposed also from a
bandstructure calculation\,\cite{saha-dasgupta2002} and is now
experimentally confirmed by our data. The smaller $J'$ arises
most probably from a weaker overlap via the small [Ar] shell of
K$^+$ than via Tl$^+$ with the configuration
[Xe]$4f^{14}5d^{10}6s^2$.

Although it is, in general, difficult to predict the microscopic
changes under (uniaxial) pressure, one may understand
qualitatively the uniaxial pressure dependencies of $J$ in a
simple microscopic picture. The dimers are formed by the
Cu$^{2+}$ spins of two neighboring CuCl$_6$ octahedra, which are
connected via a common edge of their basal planes. The Cu--Cl--Cu
bond angle amounts to $\simeq 96^\circ$. Thus the weak
antiferromagnetic coupling of \tal\ agrees with the expectation
of the Goodenough-Kanamori-Anderson rules that the coupling
changes from weakly ferro- to strongly antiferromagnetic when the
Cu--Cl--Cu bond angle increases from $90^\circ$ to $180^\circ$.
Since the line connecting the two Cl$^{-}$ ions and the $[010]$
direction have an angle of about $29^\circ$, one may expect that
pressure along $[010]$ will shorten the Cl--Cl distance. This
would increase the Cu--Cl--Cu bond angle and enhance $J$. The
opposite may be expected for pressure along $[201]$, since this
direction has an angle of about $25^\circ$ with the connection of
the Cu$^{2+}$ ions, and a shortening of the Cu--Cu distance would
lower the Cu--Cl--Cu bond angle and  decrease $J$. The normal of
the $(10\overline{2})$ plane is nearly perpendicular to both, the
Cu--Cu ($\simeq 82^\circ$) and the Cl--Cl line ($\simeq
77^\circ$). Thus one may expect that pressure on the
$(10\overline{2})$ plane will hardly change the Cu--Cl--Cu bond
angle, but will slightly increase both the Cu--Cu and the Cl--Cl
distance. Therefore the Cu--Cl distances will increase and $J$
decreases, since the overlap between the Cu-3d and Cl-2p orbitals
becomes weaker.

In summary, we have presented high-resolution measurements of
thermal expansion and magnetostriction perpendicular to the
$(010)$ and $(10\overline{2})$ planes of \tal . For both
directions the field-induced N\'{e}el order causes very pronounced
anomalies, which allow a detailed determination of the phase
boundary. There is essentially no hysteresis as expected for a
second-order phase transition.
The anomalies signal huge uniaxial pressure dependencies of the
phase boundary, e.g.\ $\pm 185$\,\%/GPa for the spin gap obtained
from $H_c(0\,{\rm K})$ with the signs depending on the direction
of pressure. Large uniaxial pressure dependencies of opposite
signs are also present for the susceptibility around 40\,K. Our
analysis unambiguously reveals that the huge pressure
dependencies of $\Delta$ arise from pressure-dependent changes of
the intradimer coupling, whereas changes of the interdimer
coupling play a minor role. Thus, the smaller interdimer coupling
in \kal\ compared to \tal\ is clearly not a consequence of
chemical pressure.

We acknowledge fruitful discussions with M.\,Gr\"{u}ninger,
A.\,Freimuth, D.\,Khomskii, M.\,Kriener, and A.\,Rosch. This work
was supported by the Deutsche Forschungsgemeinschaft through
SFB~608.


\begin{thebibliography}{10}
\bibitem{accident}
Since $\chi_{max}$ and $\Delta$ depend on two parameters $J$ and
$J'$, the
  anisotropies of $\frac{\partial \chi_{max}}{\partial p_{i}}$ and of
  $\frac{\partial \Delta}{\partial p_{i}}$ will be identical in two cases: (i)
  if pressure changes only one parameter (in our case $J$) or (ii) if the
  anisotropies of $\frac{\partial J}{\partial p_{i}}$ and of $\frac{\partial
  J'}{\partial p_{i}}$ are identical. Since the second case would be
  accidental, it appears very unlikely.
\end{thebibliography}

\end{document}